# The Systems of Computer Mathematics in the Cloud-Based Learning Environment of Educational Institutions

Mariya Shyshkina[1], Ulyana Kohut[2], Maya Popel[1]

[1]Institute of Information Technologies and Learning Tools of NAES of Ukraine,
9 M.Berlynskoho St., Kyiv, Ukraine
`{shyshkina, popel}@iitlt.gov.ua`
[2]Drohobych Ivan Franko State Pedagogical University, 24 I.Franko Str., Drogobych
`(ulyana_kogut@mail.ru)`

**Abstract.** The article highlights the promising ways of providing access to the mathematical software in higher educational institutions. It is emphasized that the cloud computing services implementation is the actual trend of modern ICT pedagogical systems development. The analysis and evaluation of existing experience and educational research of different types of mathematical software packages use are proposed. The methodological guidance on the organization of the cloud-based learning environment in higher educational institutions using the systems of computer mathematics (SCM) is outlined. The advantages and disadvantages of different cloud service models of access to SCM are described. The cloud-based learning component with the use of Maxima system is described and evaluated. .

**Key Terms.** Methodology, InformationCommunicationTechnology, ICTTool, Educational Process.

## 1 Introduction

### 1.1 Research Objectives

Nowadays, the cloud computing is among the leading technological trends in the formation of information society. This is the core of the innovative concepts of learning and their implementations make significant impact on the content and forms of education [2; 13; 16].
Cloud computing (CC) technology is used to enhance multiple access and joint use of educational resources at different levels and domains, combining the corporate resources of the university and other learning resources within a united framework. Progress in the area has provided new insights into the problems of educational electronic resources provision and configuration within the learning environment, bring-ing new models and approaches. These tools make the great impact of the data processing changing the content, methods and organizational forms of learning, lifting the restrictions or significantly improving the access for all participants.

A separate set of problems concerns to the application of software packages for the implementation of various mathematical operations, actions and calculations, these are the so-called Systems of Computer Mathematics (SCM), including Maple Net, MATLAB web-server, WebMathematica, Calculation Laboratory and others [13; 19; 8]. This is one of the most common types of mathematical software, which is a part of the modern educational environment of educational institutions [2]. The problems emerge when searching for promising ways and models to use this type of cloud-based tools being an essential factor of engineering and mathematics disciplines training quality improving.

**The aim of the article** is to analyze the state of the art of mathematical software use and design within the cloud-based settings and evaluation of the cloud-based learning component with the use of Maxima system.

### 1.2 The Problem Statement

According to recent studies [2; 10; 16; 3] the challenges of making the ICT-based learning components of the university environment fit the needs of its users, have led to the search for the most reasonable ways for their design and delivery within the cloud-based settings. The cloud-based components possess many progressive features including better adaptability and mobility, as well as full-scale interactivity, free network access, a unified structure among others [2, 10, 12]. So, the modeling and analysis of their design and deployment in view of the current tendencies of modern advance of the cloud-based mathematical software and available learning experience have come to the fore.

Among the priority issues there are those concerning the existing approaches and models for electronic educational resources delivery within the different cloud-based service models; the cloud-based learning components elaboration, assessment and testing; research indicators substantiation.

### 1.3 The Research Methods

The research method involved analysing the current research (including the domestic and foreign experience of the cloud-based learning services and mathematical software use in educational institutions in Ukraine and abroad), evaluation of existing approaches to software delivery, their advantages and disadvantages; comparison of promising ways of popular mathematical software implementation "in the cloud", examining of models and approaches, technological solutions and psychological and pedagogical assumptions about better ways of introducing innovative technology into the learning process. The cloud-based component with the use of Maxima system was designed and elaborated within the study undertaken in 2012-2014 in the Institute of Information Technologies and Learning Tools of NAES of Ukraine devoted to the use of the SCM for informatics bachelors training (U.Kohut). The special indicators to reveal ICT competence of educational personnel trained within the cloud-based learning environment and also the learning components quality evaluation indicators were elaborated within the research work devoted to the university cloud-based learning and research environment formation and development held in 2012-2014 in the Institute of Information Technologies and Learning Tools of NAES of Ukraine (M.Shyshkina). To measure the efficiency of the proposed approach the pedagogical experiment was undertaken in Drohobych Ivan Franko State Pedagogical University. The expert quality evaluation of the CC-based components elaborated in the study was implemented. The approach and methodology were grounded within the research work "Methodology of the cloud-based learning environment of educational institution formation" that was held in the Institute of Information Technologies and Learning Tools of NAES of Ukraine in 2015-2017, Registration number 0115U002231 (coordinated by M.Shyshkina).

## 2    The State of the Art

According to the recent research [2; 5; 7; 15; 11; 14], the problems of cloud technologies implementing in educational institutions so as to provide software access, support collaborative learning, research and educational activities, exchange experience and also project development are especially challenging. The formation of the cloud-based learning environment is recognized as a priority by the international educational community [9], and is now being intensively developed in different areas of education, including mathematics and engineering [1; 4; 16; 18].

The transformation of the modern educational environment of the university by the use of the cloud-based services and cloud computing delivery platforms is an important trend in research. The topics of software virtualization and unified university ICT infrastructure formation on the basis of CC have become increasingly popular lines of investigation [8, 21]. The problems with the use of private and public cloud services, their advantages and disadvantages, perspectives on their application, and targets and implementation strategies are within the spectrum of this research [3; 4; 16].

There is a gradual shift towards the outsourcing of ICT services that are likely to provide more flexible, powerful and high-quality educational services and resources [2]. There is a tendency towards the increasing use of the software-as-a-service (SaaS) tool. Along with SaaS the network design and operation, security operations, desktop computing support, datacentre provision and other services are increasingly being outsourced as well. Indeed, the use of the outsourcing mechanism for a non-core activity of any organization, as the recent surveys have observed happening in business, is now being extended into the education sector [5]. So, the study of the best practices in the use of cloud services in an educational environment, the analysis and evaluation of possible ways of development, and service quality estimation in this context have to be considered.

The valuable experience of the Massachusetts institute of technology (MIT) should be noted in concern to the cloud based learning environment formation in particular as for access to mathematical software. The Math software is available in the corporate cloud of the University for the most popular packages such as Mathematica, Mathlab, Maple, R, Maxima [19]. This software is delivered in the distributed mode on-line through the corporate access point. This is to save on license pay and also on computing facilities. The mathematics applications require powerful processing so it is advisable to use it in the cloud. On the other case the market need in such tools inspires its supply by the SaaS model. This is evidenced by the emergence of the cloud versions for such products as SageMathCloud, Maple Net, MATLAB web-server, Web-Mathematica, Calculation Laboratory and others [1; 4]. Really there is a shift toward the cloud-based models as from the side of educational and scientific community, and also from the side of product suppliers. The learning software actually becomes a service in any case; let it be a public or a corporate cloud [12].

An essential feature of the cloud computing conception is dynamic supply of computing resources, software and hardware its flexible configuration according to user needs. So comparison of different approaches and cloud models of software access is the current subject matter of educational research [3; 4; 11; 17]. Despite of the fact that the sphere of CC is rather emerging there is a need of some comparison of the achieved experience to consider future prospects [17]. Also the problems of software choice in the learning complexes to be implemented in a cloud arise. This leads to the problems of cloud-based learning resources evaluation techniques and research indicators substantiation.

The special attention is to be paid to the system Maxima, because it is easy to master, it is comparable to such systems as Maple and Mahtematisa as for solving problems (for example in the field of operations research) and it is free accessed. It is equipped with a menu system that enables characters conversion, solve equations, compute derivatives, integrals, etc., avoiding some additional efforts as for learning the special language tools to implement these actions. In view of that the Maxima system can be used to study Math and Computer Science disciplines even on the first year of study at the pedagogical university [13]. The use of Maxima will not cause any difficulties for the students as for solving problems of mathematical analysis and linear algebra – the students are required only to choose a menu item and enter the expression. However, programming within Maxima requires knowledge of certain language and syntax, as well as some commands.

Thus, in view of the current tendencies, the research questions are: how can we take maximum advantage of modern network technologies and compose the tools and services of the learning environment to achieve better results? What are the best ways to access electronic resources and mathematical software if the environment is designed mainly and essentially on the basis of CC? What are the most reasonable approaches to evaluate the results? This brings the problem of the cloud-based learning components modelling, evaluation and design to the forefront.

## 3  The Research Results

With the development of cloud-based network tools and technologies new kinds of services and applications emerge that may be used to support math and engineering disciplines learning. There is a shift towards greater use of on-line network tools in educational universities, among them there are such as:
- platforms of distance learning support (Moodle, LearningSpace, Sakai, Blackboard, etc.), including online resources (Competentum.ONLINE, Google Open Class, Canvas) [10];
- mathematical software for special purposes - for example, Maple Net, MATLAB web-server, WebMathematica, Calculation Laboratory and others [13; 17; 19].

An The use of these technologies adds and provides an opportunity to explore and develop new approaches to learning, which in turn leads to the development of new strategies and methodology of teaching of mathematics disciplines in educational universities [13].

A separate set of problems relates to the use of mathematical software tools within the cloud-based learning environment. This raises the possibility of computer capacities, software and hardware dynamic delivery and its flexible configuration due to the user needs. With this approach, the organized access to various types of electronic educational resources may be arranged being specially installed on a cloud server and provided as a public service. [2]

"On this basis the subject-technological organization of information learning space, organized processes of accumulation and storage of various domain collections of electronic educational resources ensures equal access to them for learners, significantly improving ICT support of learning, research and educational management " [2, with. 11].

By means of the cloud services the applications may become available to users, as well as storage space and also computer capacities [2]. The main types of cloud service models reflect the possible ways of ICT outsourcing use in educational environment of the university in particular there are SaaS (Software-as a Service) - "software as a service", PaaS (Platform as a Service) - "platform as a service", IaaS (Infrastructure as a Service) [2, 3, 4, 9].

Recently many software applications and packages of mathematical destination started the cloud versions supplied by SaaS model, which can be used in the learning process and research as specialized software [12]. In this case, organized access to ready-made software is supported on the vendor server.

For example, the Sage mathematical software is supplied by this model. The system is designed to operate Sage and support experimentation with algebraic and geometric objects. As open source software, that can be used to take advantage of a variety of packages for operations on mathematical analysis, algebra, group theory, graph theory, and others. Currently, the cloud versions of SageMatCloud can do it directly from the browser. Now this is a freely available service supported by the server of the University of Washington. There is a cluster for supporting SageMathCloud, containing 288 cores, 1.2TB RAM and 50TB of storage.

In turn, the IaaS model is designed to run any application on the cloud hardware of the provider configured and selected by the user. The IaaS composition may include hardware (servers, storage, client systems and equipment); operating systems and software (virtualization, resource management); software communication tools (network integration, resource management, equipment management) provided over the Internet [12].

Using this technology there is no need to maintain complex infrastructures, data processing, network applications and client server organization, but yet renting them as a service. Specifically, users can get, at their disposal, a completely prepared virtualized workplace. This raises the possibility to provide significant amount of educational content by means of quite cheap hardware (this may be a laptop, a netbook or even a smartphone) [2].

Recently there has been a trend towards convergence and integration of various mathematical packages. For example, the latest versions of Mathematica and Maple are supplied by powerful tools for visual programming; MathCAD can work together with MatLab etc. So, for the aim of practical training any of the above packages may be used with regard to specific traditions and support opportunities of educational institutions [13]. These factors significantly influence the choice of software that can be installed "in the cloud".

Given the factors of the license application and other features the Maxima system may be advised to be used, because:

— the system is distributed under the license GNU / GPL;
— it is equipped with a system of menu that has a Ukrainian-language interface;
— it is one of the best on the implementation of symbolic computation (in fact, the only one that can compete with commercial Maple and Mathematica) [13].

Recently Given the existence of different models of cloud services use the special attention to a balanced selection of the most appropriate solutions that fit each case for a particular organization, both collective and individual user should be made. The SaaS model choice in this respect could be justified by the fact that these services are the most affordable to use, even though they need a thorough analysis of the market and educationally prudent selection of software application, to achieve the desired educational or research purposes. These tools can be involved in the learning process by individual teacher, on the level of a department, for individual or collective users.

At the same time, the construction of the ICT infrastructure of an institution as a whole needs to select and analyze the appropriate cloud platform that can be organized by the models of PaaS or IaaS. This requires the solution of a number of organizational issues, such as the formation of a special ICT unit of employees that are to be qualified to configure and deploy that infrastructure, the construction of the necessary hardware and software, the determining the plan and design phases, the designing and test of information and educational environment, filling it with the necessary resources to implement and monitor their quality, training teaching staff etc. [2]. In this case, given the results of foreign experience and current trends of the IT sector development, we can conclude that the most appropriate is the use of hybrid service models that can be incorporated by means of public and corporate clouds, not excluding the means of the model of "software as a service" if necessary. [2]

Due to the possibilities of resources sharing, the cloud services provide a base for easy access to educational services, in line with the principles of open education, combining of science and practice, integrating the processes of training and scientific research.

### 3.1 The Design of the Cloud-based Learning Component with the Use of the Maxima System

To research the hybrid service model of learning software access, especially for the mathematical software delivery a joint investigation was undertaken in 2013–2014 at the Institute of Information Technologies and Learning Tools of the NAES of Ukraine and Drohobych State Pedagogical University named after I.Franko. At the pedagogical experiment the cloud-based learning component with Maxima system was designed and used for the operations research study.

In this case, the implementation of software access due to the hybrid cloud deployment was organised.

The configuration of the virtual hybrid cloud used in the pedagogical experiment was described in [12]. The model contains a virtual corporate (private) subnet and a public subnet. The public subnet can be accessed by a user through the remote desktop protocol (RDP). In this case, a user (student) refers to certain electronic resources and a computing capacity set on a virtual machine of the cloud server from any device, anywhere and at any time, using the Internet connection.

The advantage of the proposed model is that, in a learning process, it is necessary to use both corporate and public learning resources for special purposes. In particular, the corporate cloud contains limited access software; this may be due to the copyright being owned by an author, or the use of licensed software products, personal data and other information of corporate use. In addition, there is a considerable saving of computational resources, as the software used in the distributed mode does not require direct Internet access for each student. At the same time, there is a possibility of placing some public resources on a virtual server so the learner can access them via the Internet and use the server with the powerful processing capabilities in any place and at any time. These resources are in the public cloud and can be supplied as needed [12].

Within the experimental study the Maxima system installed on a virtual server running Ubuntu 10.04 (Lucid Lynks), was implemented. In the repository of this operational system is a version of Maxima based on the editor Emacs, which was installed on a student's virtual desktop [13].

To create a session (to insert an item Maxima) you should choose the menu option Insert - Session - Maxima. There is an active input line to input Maxima commands.

An example of successful use of the Maxima cloud-based component is Graph Theory learning. Maxima has a rich set of features on the design and elaboration of relevant objects of this theory. Some examples of its use are presented at [12].

## 3.2 The Cloud-based Component Implementation and Evaluation

In the research experiment held at Drohobych State Pedagogical University named after I.Franko, 240 students participated. The aim was to test the specially designed learning environment for training the operations research skills on the basis of Maxima system. During the study, the formation of students' professional competence by means of a special training method was examined. The experiment confirmed the rise of the student competence, which was shown using the $\chi^2$–Pearson criterion [13]. This result was achieved through a deepening of the research component of training. The experiment was designed using a local version of the Maxima system installed on a student's desktop.

The special aspect of the study was the expansion of these results using the cloud version of the Maxima system that was posted on a virtual desktop. In the first case study (with the local version), this tool was applied only in special training situations. In the second case study (the cloud version), the students' research activity with the system extended beyond the classroom time. This, in turn, was used to improve the learning outcomes.

At the next stage of the experiment (2014-2015) the aim was to test the use of the cloud-based component in the learning process. 48 students participated in this experiment. There was the experimental group of 24 students who used the cloud-based component with Maxima system. It showed increase of the students' percentage with the high level of ICT competence from 16% to 75%. It was significantly different from the level of ICT competence of those students who did not used this component (from 14% to 20%), which was justified by the Fisher criterion.

The cloud-based learning component used in the experiment has undergone a quality estimation. The method of learning resources quality estimation developed in the joint laboratory of educational quality management with the use of ICT [6] was used and adapted for this study. The 20 experts were specially selected as having experience in teaching professional disciplines focused on the use of ICT and being involved in the evaluation process. The experts evaluated the electronic resource by two groups of parameters. The first group has contained 7 technological parameters: ease of access; the clarity of the interface; sustainability; support of collaborative work, ease of integration; mobility; and usefulness. The second group has contained 9 psychological and pedagogical parameters: the scientific clarity; accessibility; fostering the intellectual development; problem orientation; personalization; adaptability; methodical usefulness; professional orientation; and feedback connection.

The problem was: is it reasonable and feasible to arrange the environment in a proposed way? For this purpose there were two questionnaires proposed to expert concerning two groups of parameters. The 20 experts estimated 16 parameters (there were 7 technological and 9 psychological and pedagogical among them). A four-point scale (0 (no), 1 (low), 2 (good), 3 (excellent)) was used for the questions.
The resulting average value was calculated for every parameter among the technological ones : "Ease of access" = 2.1, "Interface clarity" = 2.4, "Responsiveness" = 2.1, "Sustainability" = 2.56, "Support of Collaborative work" = 2.0, "Ease of Integration" = 2.0, "Usefulness" = 2.8, the total value was 2.3.

The resulting average values for every psychological and pedagogical parameter was calculated as: "Scientific clarity" = 2.6, "Accessibility" = 2.7, "Fostering the intellectual development" = 2.5, "Problem orientation" = 2.8, "Personalization" = 2.8, "Adaptability" = 2.6, "Methodical usefulness" = 2.81, "Professional orientation" = 2,75, "Feedback connection" = 2,75. The total value was 2.71.

The resulted average criterion of EER quality K=2,59. This characterises the resource quality as sufficient for further implementation and use.

The advantage of the approach is the possibility to compare the different ways to implement resources with regard to the learning infrastructure. Future research in this area should consider different types of resources and environments.

## 4      Conclusions and Prospects for Further Research

The results of the study indicate certain movement in the development of new ways to create and use the software for educational purposes based on the concept of cloud computing.

The methodological guidelines on the organization of a cloud-based learning environment of computer science and mathematics disciplines cover a range of delivery services and software applications to be available for use. Choosing the SaaS model in this respect could be justified by the fact that these services are the most affordable to use, regarding the restriction in the choice of software application to those offered by a supplier. When choosing a PaaS or IaaS model one must take into account a number of technological and organizational factors forming the environment, and quality characteristics of software selection which need to be installed in the cloud.

The introduction and design of the cloud-based learning components into the process of math and computer science study contributes to the growth of access to the best examples of electronic resources and services, ICT tools modernization, better learning outcomes. The cloud-based learning component on the basis of Maxima system was successfully developed and justified within these settings. Its use is promising regarding learning design that can account for the advances and tendencies of CC progress.